\documentclass[twocolumn,showpacs,preprintnumbers,amsmath,amssymb]{revtex4}

\usepackage{graphicx}                            
\usepackage{dcolumn}                             
\usepackage{bm}                                  
\usepackage{verbatim}

\def\be{\begin{equation}}
\def\en{\end{equation}}

\begin{document}

\title{Onion structures 
induced by\\ 
hydrophilic and hydrophobic ions 
 in a binary  mixture}
\author{Koichiro Sadakane}
\affiliation{Department of Physics, Graduate School of Science, Kyoto University, Kyoto 606-8502, Japan}
\author{Akira Onuki}
\affiliation{Department of Physics, Graduate School of Science, Kyoto University, Kyoto 606-8502, Japan}
\author{Koji Nishida}
\affiliation{Institute for Chemical Research, Kyoto University, Uji, Kyoto 611-0011, Japan}
\author{Satoshi Koizumi}
\affiliation{Advanced Science Research Center, Japan Atomic Energy Agency, Tokai, 319-1195, Japan}
\author{Hideki Seto}
\email{hideki.seto@kek.jp}
\affiliation{Institute of Materials Structure Science, High Energy Accelerator Research Organization, Tsukuba 305-0801, Japan}

\date{\today} 

\begin{abstract} 
Phase transition is observed 
between  one-phase disordered phase 
and  an ordered phase with 
 multi-lamellar (onion) structures 
in  an off-critical
 mixture of D$_2$O and 3-methylpyridine (3MP) containing 
a salt at 85mM. 
The salt consists  of hydrophilic cations 
and hydrophobic anions, which interact asymmetrically 
with the solvent composition fluctuations  inducing  
 mesophases. The structure factor of the  composition distribution
 obtained  from   small-angle neutron scattering 
has a peak at an intermediate wave number  in the disordered  
phase  and multiple peaks in the ordered  
phase.   Lamellar layers forming onions are   
 composed of solvation-induced charged 
membranes swollen by D$_2$O. 
The onion phase is  realized   only for small 
 volume fractions  of 3MP (in D$_2$O-rich solvent).
\end{abstract}
\maketitle


Binary mixtures of water and organic solvent have been 
used extensively    to study  universal aspects of 
  critical behavior and phase separation dynamics. 
However,  not enough attention  
  has yet  been paid  to unique ion 
 effects in  such  mixtures,  where  
 preferential hydration around each ion 
should affect the composition fluctuations 
\cite{Is,Marcus}. 
Salts composed of small 
hydrophilic  ions  can  
drastically change  the phase behavior 
 even  at small concentrations \cite{Eckfeldt, Hales, Balevicius, Misawa, Takamuku}.  
More strikingly,  many  
groups have observed long-lived 
heterogeneities (sometimes extending  over 
a few micrometers) 
 in one-phase state \cite{So,Wagner}
and a  third phase 
visible  as a thin plate formed 
at  a liquid-liquid interface in two-phase state \cite{third}. 
These   observations were  reproduced  in different experiments 
using  solvents and salts of  high  purity, so they 
 should  be  regarded as ion-induced supramolecular 
aggregates.  We also comment on 
 a mixture of water+isobutylic acid (IA), where 
IA  partly
dissociates into H$^+$ and Butyrate$^-$   ions. 
There, the   mobility of  H$^+$ ions  much decreases       
at large IA contents  in  one-phase state  \cite{Bonn} 
and the third phase 
also appears    around an interface  
in  two-phase state \cite{third}. 
Similar phenomena have often been 
observed in various soft matters 
such as polymers, gels, colloids, 
and mixtures containing  surfactants. 
Though not well understood, 
the solvation (ion-dipole) interaction among 
ions and polar  molecules should
play a major role in these phenomena 
together with the Coulombic 
interaction among charges. \cite{Is,Marcus}

Recently, the solvation effect 
on phase behavior was theoretically studied 
in polar mixtures including the case of 
 antagonistic ion pairs 
composed  of hydrophilic and hydrophobic ions 
\cite{OnukiJCP04}. 
Such cations and   anions interact    
differently   with the 
composition   fluctuations, 
leading to  a charge-density-wave phase 
for sufficiently large solvation asymmetry 
 even at small 
salt concentrations. 
In a small-angle neutron scattering (SANS) 
experiment,  Sadakane 
{\it et al.} \cite{SadakaneJPSJ07} found  a  peak   
at an intermediate wave number $Q_m$($\sim 0.1~$\AA$^{-1}$)     
 in a    near-critical mixture of D$_2$O 
and 3-methylpyridine (3MP) 
containing  sodium tetraphenylborate 
NaBPh$_\mathrm{4}$  at $ 100$ mM. 
The  volume fraction of 3MP, denoted by 
$\phi_{3{\rm MP}}$, was chosen to be 
0.35, 0.42,  and 0.54. 
This  salt  dissociates   
into  hydrophilic  Na$^+$ and 
 hydrophobic  BPh$_\mathrm{4}$$^-$. The BPh$_\mathrm{4}$$^-$  is   
composed of  four phenyl rings bonded to an ionized 
boron.   This anion   acquires  strong hydrophobicity  
such that the  salt is   more soluble to pure  3MP 
than to pure D$_2$O  despite the hydrophilic nature 
of  Na$^+$  \cite{note}.  Furthermore,  
the mixture exhibited colors 
changing dramatically on approaching 
the criticality at low salt concentrations $(\sim 10~$mM), 
indicating emergence     of large scale heterogeneities. 
No distinct mesophases  appeared in the same solvent 
when   hydrophilic salts  such as NaCl were added 
\cite{SadakaneCPL06}.

In this Letter, we again use the system of D$_2$O+ 3MP+NaBPh$_{4}$.  
We purchased D$_2$O with isotopic purity of 99\%  
from EURISO-TOP and   99\%  3MP and 
99.5 \%   NaBPh$_{4}$ from Aldrich. These chemicals were 
mixed without further purification. 
As a new finding, we report observation of  
  multi-lamellar (onion) structures 
for small $\phi_{3{\rm MP}}$.  
 Detailed measurements were made for  
 $\phi_{3{\rm MP}}=$0.08, 0.09, 0.11, and 0.14
in the temperature range $283$~K $ \le T \le 343$~K. 
The concentration of NaBPh$_4$ was 
fixed at 85 mM.  In this mixture solvent, 
there is  a lower  critical 
solution temperature (LCST) 
at 310 K for $\phi_{3{\rm MP}}=0.30$. 
Without salt, the mixtures of the same composition as our systems are homogeneous in this temperature range, 
and the composition fluctuation is very small. 
However, with addition of  NaBPh$_\mathrm{4}$, 
onions appeared spontaneously directly from 
disordered states. 
If $\phi_{3{\rm MP}}$ = 0.09 as an example,  
 the ratios among the  number densities 
 of D$_{2}$O, 3MP, and NaBPh$_{4}$, 
denoted by $n_{\rm D}$, $n_{3{\rm MP}}$, and $n_{\rm salt}$, 
are given by  
\be 
n_{\rm D}/n_{3{\rm MP}}
\cong 54.3, \quad 
n_{3{\rm MP}}/n_{\rm salt} \cong 10.9.
\label{nD}
\en 
Thus $n_{\rm salt }\ll n_{3{\rm MP}}\ll n_{{\rm D}}$. 
The molecular volumes of 
 D$_{2}$O and 3MP (the inverse densities of 
the pure components) are 28 and 
 168 \AA$^3$, respectively.  The volume fraction of NaBPh$_{4}$ was  
less than 2 $\%$  and the observed SANS intensity 
was mostly due to  the composition distribution   
of the solvent. Also in surfactant mixtures, 
the spontaneous onion formation (without applying shear flow) 
 has  been observed 
at small surfactant concentrations in water-rich solvent  
\cite{Berg,Ramos,Herve,Iwashita}.

\begin{figure}[htbp]
  \begin{center}
    \includegraphics[width=7.2cm,height=8.57cm]{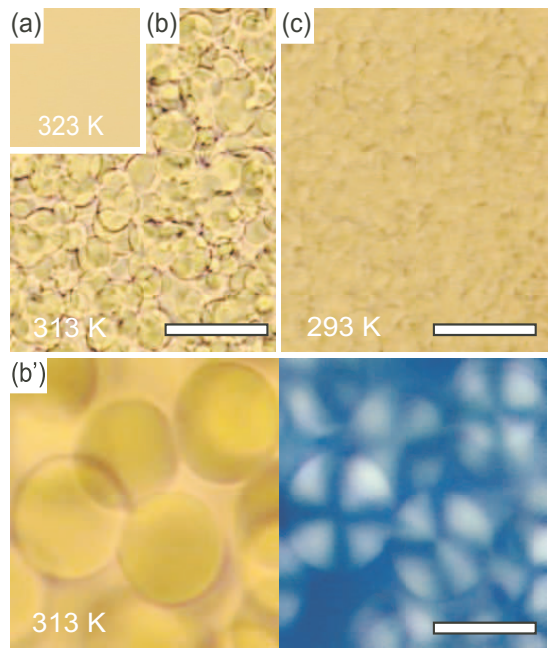}
  \end{center}
\caption{
Optical microscopic images obtained from the sample of $\phi_{3{\rm MP}}$ = 0.09;  
(a) $T=323$ K (disordered), 
(b) $T=313$ K (onion), and 
(c) $T=293$ K (onion). 
Scale bars in (b) and (c) 
indicate 100 $\mu$m. 
(b$^{\prime}$) A magnified image of (b) (left) 
and a birefringent    image of  the same region  
(right) with  scale bar being  20 $\mu$m. 
}
\label{MO}
\end{figure}

Figure \ref{MO} gives temperature dependence of 
optical microscopic images. These 
images were taken using a Nikon Optiphot2-Pol with a CCD camera.  The thickness of the sample was 90~$\mu$m and the temperature of the samples was controlled within an accuracy better than 0.1 K using a Linkam TH-99 hot stage.
The composition of the sample was   
$\phi_{3{\rm MP}}$ = 0.09. 
At $T=323~$ K (Fig. \ref{MO} (a)), the whole area 
is homogeneous without visible structures. 
As we decreased  $T$  below $318~$K, 
small droplets  emerged from the whole sample and
 growed  in size. (Fig. \ref{MO} (b)). 
In (c),  $T$ is further lowered to $293~$K  
and  the space is nearly 
filled with  deformed droplets with  diameter 
 about $20~ \mu$m. 
The left panel  of (b$^{\prime}$) 
gives  a magnified picture of (b), while 
the right panel is the 
corresponding   Maltese cross pattern obtained by 
crossed Nicoles  under polarized light. 
 This  pattern arises from inhomogeneous composition 
distributions with spherical symmetry. Such patterns 
 have been observed for    
polymer spherulites  \cite{Stein} 
and  onions of surfactant systems 
\cite{Berg,Ramos,Iwashita}.
Thus our  result indicates  that 
the onion structure is 
formed below $T_L$, 
where  $T_L = 318~$K is the transition temperature.  


\begin{figure}[htbp]
  \begin{center}
    \includegraphics[width=6.5cm]{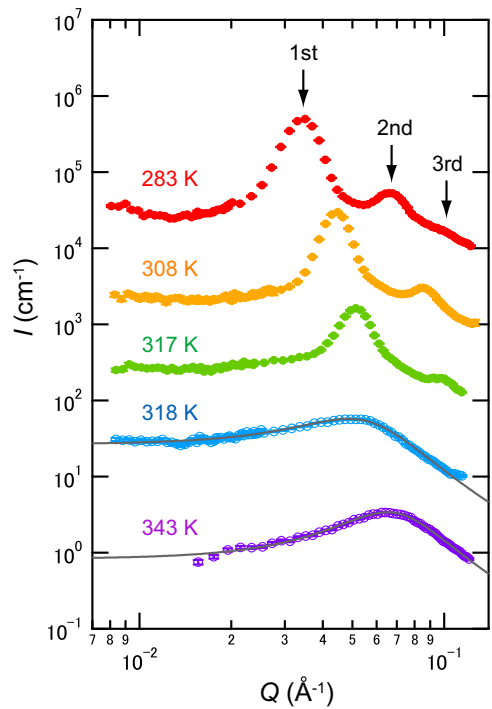}
  \end{center}
\caption{
Temperature dependence of SANS intensity 
from a sample of $\phi_{3{\rm MP}}$ = 0.09.
First-order transition occurs slightly below  $T=318~$K.  
Data at lower temperatures are 
shifted by multiplying 10 from below.}
\label{SANS}
\end{figure}

SANS  experiments were performed at SANS-J, 
JRR-3M in Japan Atomic Energy Agency.
A 6.0 \AA ~incident neutron beam was 
mechanically selected with a
wavelength resolution of 13 \%, 
and the scattered neutrons were detected
at 2.5 m and 10 m from the sample position.
Each sample was kept in a quartz cell of 2 mm thickness 
and placed in a temperature-controlled chamber 
with an accuracy better than 0.1 K.
The measured wave number $Q$ 
ranged   between $6 \times 10^{-3}~$\AA$^{-1}$ and
$1.3 \times 10^{-1}~$\AA$^{-1}$ and 
the observed two-dimensional data 
were azimuthally averaged.
Figure \ref{SANS} displays 
the  SANS intensity  vs  $Q$ 
  from the sample of $\phi_{3{\rm MP}}$ = 0.09. 
With  varying  $T$, 
 the system was in the disordered phase 
for $T>T_L$ and  in the ordered phase 
for $T<T_L$.
 At the lowest temperature $283$~K in Fig. \ref{SANS}, 
a Bragg peak is pronounced at 
$Q_m=3.4 \times 10^{-2} ~\mathrm{\AA}^{-1}$,  
 the second one appears at $6.8 \times 10^{-2}$~{\AA}$^{-1}$,  
and  a slight shoulder exists around $1.02 \times 10^{-1}~${\AA}$^{-1}$
(though not clearly seen in the figure). 
 The same behavior was 
found  for all the SANS data 
below $T_L$, demonstrating the  formation of 
 aligned lamellae.  This is another evidence of the onion formation below $T_L$.
 
\begin{figure}
  \begin{center}
\includegraphics[scale=0.9]{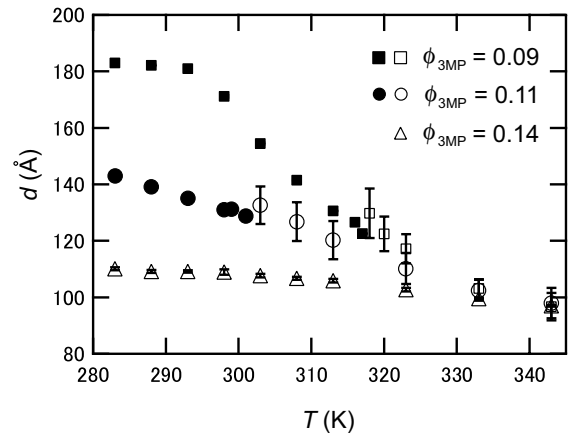}
  \end{center}
\caption{
Repeat distance $d=2\pi/Q_m$ vs temperature 
$T$ for 
 $\phi_{3{\rm MP}}$ = 0.14 (triangle), 0.11 (circle), and 0.09 (square). 
Open symbols are  data in the disordered phase, while 
 filled  symbols in the onion  phase. No onion 
appears for $\phi_{3{\rm MP}}$ = 0.14.
}
\label{temp_dep}
\end{figure}

In the disordered phase above $T_L$, the intensity 
 exhibits   a broad peak 
at $Q_m \sim 0.07~$\AA$^{-1}$,    
with $Q_m$  slightly shifting  
to lower  values as $T\to T_L$. It 
  may fairly be fitted to 
the theoretical mean-field intensity $I_{\rm{OK}}(Q)$
 (solid curves for $T=343~$ and $318~$K 
in Fig. \ref{SANS}) \cite{OnukiJCP04}. Its inverse is of the form 

\begin{equation}
{I_{\rm{OK}}(0)} /{I_{\rm{OK}}(Q)} 
= { 1+\bigg[1-\frac{\gamma_p^2}{1+{\lambda}^{2}Q^{2}}
\bigg ] {\xi}^{2}Q^{2} }
\label{Onuki}~,
\end{equation}

\noindent where $\gamma_p$  
represents  the degree of solvation asymmetry 
between cations and anions. For $\gamma_p=0$ 
it follows the Ornstein-Zernike form.  For $\gamma_p>1$
 the structure factor has a peak 
at an intermediate wave number. 
In the present  case $\gamma_p$ is taken as an adjustable 
parameter given by 
2.32 at $T=333$~K and  by 1.96 at $T=320$~K. 
Slightly below $T_L$,  the system 
was phase-separated into 
an  ordered phase in a  upper region 
and the disordered phase in a lower  region. 
The data in the case 
 $T<T_L$ in Fig. \ref{SANS} are those from the upper region. 
The transition is thus first-order. 
In addition,  the same $T_L$ was obtained with decreasing 
and increasing  $T$ around the transition. 
The mass density  of 
D$_{2}$O  is $1.11 ~$g$/$cm$^3$ 
and that of  3MP is 
$0.96 ~$g$/$cm$^3$, so  the 
lighter ordered phase contains more 3MP than the coexisting 
disordered phase.  In this case  
the mass density difference between the two phases 
is less than  $10$~mg$/$cm$^3$. 
As $T$ was  further lowered, the onion droplets  
became more swollen by D$_2$O. This results  in   
an increase of the lamellar spacing, as 
will  be shown   in Fig.  \ref{temp_dep}.   
At low temperatures 
below  $ 293$~K,  the whole 
cell  was filled with swollen 
onions.

The lamellar/disorder transition 
 occurred only in 
a window  composition range 
 $0.05 < \phi_{3{\rm MP}}< 0.12$.
Figure \ref{temp_dep} shows the temperature dependence of 
the mean repeat distance $d=2\pi/Q_m$  
calculated from the peak position $Q_m$.  
For  $\phi_{3{\rm MP}}=0.09$,  $d$ considerably 
increases  with decreasing $T$ 
in the range $293~$K$<T< T_L$, which is due to the  swelling of  
the onions.  The relaxation time 
of this swelling was shorter  than 1 min. 
For $\phi_{3{\rm MP}}=0.14$, 
 $d$ varies very weakly in the whole temperature range.  
Above $330$ K, $\phi_{3{\rm MP}}$ dependence of $d$ is weak.  
The  transition temperature 
$T_L$ decreases with increasing $\phi_{3{\rm MP}}$, 
being equal to  $318$~K and $303$~K 
for  $\phi_{3{\rm MP}} =0.09$ and $0.11$, respectively. 
However, the system remained in the disordered 
phase for $\phi_{3{\rm MP}}\ge 0.14$. 
The figure indicates  that 
the transition occurs for  $d\sim 130$ \AA.

\begin{figure}[htbp]
  \begin{center}
\includegraphics[scale=0.9]{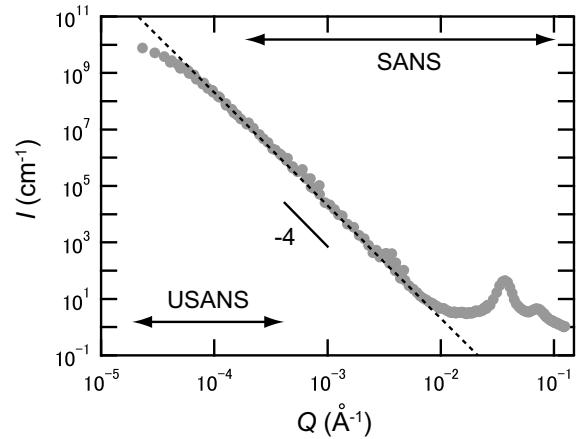}
  \end{center}
\caption{ Connected USANS  profile obtained at PNO and SANS-J
over a wide range of  $Q$  at $T=298$K and 
 $\phi_{3{\rm MP}}$ = 0.09. 
The Porod tail $\propto Q^{-4}$ from onion surfaces 
can be seen in the range  $1\times 
10^{-4}<Q<5 \times 10^{-3}$\AA$^{-1}$.
}
\label{USANS}
\end{figure}

Figure \ref{USANS} shows 
a ultra small-angle neutron scattering (USANS) spectrum observed  at PNO spectrometer in JRR-3M ($2 \times
10^{-5}<Q<5 \times 10^{-4}${\AA}$^{-1}$)
together with the data obtained at SANS-J in focusing-SANS mode (the lowest-$Q$ is expanded down to $3 \times 10^{-4}$ \AA$^{-1}$ \cite{Koizumi}). 
The measurement was done at 
$T=298$~K for the sample of $\phi_{3{\rm MP}}=0.09$.
 The low-$Q$  profile obeys 
 the Porod  law in the range of
 $1 \times 10^{-4} < Q < 5 \times 10^{-3} ${\AA}$^{-1}$. 
It should arise from the interfaces 
of the onion droplets in the range 
$R^{-1} \ll Q\ll Q_m $, where $R$ is the droplet radius. 
Each onion is composed of about 1000 concentric lamellae provided if the droplet interior is filled with layers.

Now we discuss the physical processes involved. 
In aqueous solutions,  each   Na$^+$ ion 
is known to  be surrounded by  a hydration shell composed of 
several water molecules \cite{Is,Marcus}. 
This then suggests that each 
BPh$_\mathrm{4}^-$ ion 
should   be solvated by 
  a certain number of 3MP molecules around it 
due to its strong hydrophobicity \cite{note}, 
though we do not know this solvation  number $N_{3{\rm MP}}^s$ 
at present. 
See  eq. (\ref{nD}) for the average 
number densities of BPh$_\mathrm{4}^-$ and 
3MP, $n_{\rm salt}$ and $n_{3{\rm MP}}$, as typical values 
 in the disordered phase at  $\phi_{3{\rm MP}}=0.09$. 
Then, the  number density of the solvating 3MP 
molecules is  $N_{3{\rm MP}}^s n_{\rm salt}$, 
while that of the  free  3MP 
molecules is  $n_{3{\rm MP}}- N_{3{\rm MP}}^s n_{\rm salt}$. 
In the onion phase, 
each onion should be  composed of concentric 
thin membranes made of BPh$_\mathrm{4}^-$ ions 
and solvating    3MP molecules. 
If all the BPh$_\mathrm{4}^-$ ions 
are trapped on the membranes, the areal density 
of BPh$_\mathrm{4}^-$  is given by 

\be 
\sigma_a= n_sd= 92 ~({\rm \AA}^2). 
\en 

\noindent Here  $d=180$~\AA, which is the spacing  in  
the case  (c) in Fig. \ref{MO}. On both sides of  each membrane,  
Na$^+$ ions are localized within ``counterion layers"  
with thickness of the order of the Debye screening length 
$\lambda = (4\pi n_{\rm salt} \ell_B)^{-1/2} 
\cong 15$~\AA, where $\ell_B$ is the Bjerrum length 
in D$_2$O ($\sim 7$~\AA). The charge and composition 
segregation 
in this manner should  much lower 
 the solvation free energy. 
It is worth noting that 
hydrophilic  and hydrophobic ions  
 undergo microphase separation 
around a water-oil  interface to 
lower the solvation free energy, largely 
reducing the surface tension \cite{OnukiPRE}.

We  also argue why the lamellar structure is formed 
in the window  composition range $0.05<\phi_{3{\rm MP}}<0.12$ 
in our system. 
If  $\phi_{3{\rm MP}}$ is less than the lower bound, 
the number density of  3MP molecules should be  too small to trigger  
the aggregation of  BPh$_\mathrm{4}^-$ ions. 
If  $\phi_{3{\rm MP}}$ is increased above the upper bound, 
3MP molecules would  be abundant enough outside 
the thin membrane regions. This eventually leads to 
  destruction of  the membranes with increasing 
$\phi_{3{\rm MP}}$, where  
delocalization of  BPh$_\mathrm{4}^-$ ions  can 
increase  their translational entropy 
without penalty of 
 the solvation free energy increase.

In summary, we have  realized the onion structure 
in  a D$_2$O-3MP mixture with small 3MP contents 
by adding a small amount of an 
 antagonistic  salt composed of 
hydrophilic cations and hydrophobic anions. 
In the previous experiments $[3$-$6]$, 
hydrophilic salts have mostly 
been used and a number of unsolved 
controversies have been posed.  
 As demonstrated in this Letter,  
the  effects  of  adding 
an antagonistic salt are   even 
 more spectacular.  As future experiments, 
we plan to  measure  the 
electric conductivity and 
the viscoelastic properties in salt-induced  mesophases. 
Experiments in other soft matters 
in this direction should  also  be informative.

\begin{acknowledgments}
This work was supported by Grant-in-Aid 
for Scientific Research on Priority 
Area ``Soft Matter Physics'' from the Ministry of Education, 
Culture, Sports, Science and Technology of Japan.
One of the authors (KS) was  
supported by the JSPS Fellowships (19-3802).
Thanks were due to   Dr. D. Yamaguchi at 
JAEA for  assistance in the SANS experiment.
\end{acknowledgments}

\end{document}